\documentclass[%
 rsi,
 amsmath,amssymb,
preprint,%
]{revtex4-1}

\usepackage{graphicx}
\usepackage{dcolumn}
\usepackage{bm}
\usepackage{hyperref}
\usepackage{caption}
\usepackage{subcaption}

\begin{document}

\title{Automated Estimation of Plasma Temperature and Density from Emission Spectroscopy}

\author{Todd A.~Oliver}%
\email{oliver@oden.utexas.edu}
\affiliation{Oden Institute for Computational Engineering and Sciences\\University of Texas at Austin}

\author{Craig Michoski}%
\email{michoski@sapient-a-i.com}%
\affiliation{Sapientai LLC}
\altaffiliation[Also at ]{Oden Institute for Computational Engineering and Sciences, UT-Austin}%

\author{Samuel Langendorf}
\email{samuel.langendorf@lanl.gov}
\affiliation{Los Alamos National Laboratory}

\author{Andrew LaJoie}
\email{lajoiean@lanl.gov}
\affiliation{Los Alamos National Laboratory}

\date{\today}

\begin{abstract}

This paper introduces a novel approach for automated estimation of plasma temperature and density using emission spectroscopy, integrating Bayesian inference with sophisticated physical models. We provide an in-depth examination of Bayesian methods applied to the complexities of plasma diagnostics, supported by a robust framework of physical and measurement models. Our methodology is validated through experimental observations, focusing on individual and sequential shot analyses. The results demonstrate the effectiveness of our approach in enhancing the accuracy and reliability of plasma parameter estimation, marking a significant advancement in the field of emission spectroscopy for plasma diagnostics. This study not only offers a new perspective in plasma analysis but also paves the way for further research and applications in nuclear instrumentation and related domains.
\end{abstract}

\maketitle

\section{Introduction}

Spectroscopy is a powerful and versatile analytical tool that plays an increasingly critical role in a broad array of scientific and industrial fields. For example, atomic emission spectroscopy, centered on the study of emitted light spectra from excited atoms or molecules, offers a non-invasive and highly sensitive method for the characterization and identification of various substances. From environmental monitoring, medical diagnostics, and forensic analysis, to astrophysics, material science, and industrial process control, emission spectroscopy's extensive applications underscore its importance. 

However, the quantities observed in emission spectroscopy---i.e., the intensities of light at different wavelengths---are generally not directly of interest in applications.  Instead, it is common to extract the quantities of interest (QoI's) from these observables using a model.  As an example, the simplest way to extract information from spectra is by assuming local thermodynamic equilibrium (LTE) conditions, and then leveraging the Boltzmann distribution law, where populations of distinct atomic energy levels are charted against their corresponding energies, yielding a graph whose slope bears a reciprocal relationship to the temperature of the system.  Such plots are referred to as Boltzmann plots.  However, in many application areas, such as in transient high energy laser plasma \cite{cristoforetti2013thermodynamic}, the underlying LTE assumptions do not hold, and thus are generally considered inadequate for effectively interpreting the measured spectra.  Thus, in order to accommodate for the added complexity present under non-LTE conditions (such as the presence of large local gradients), various complex kinetic models of the plasma have been developed and deployed, which effectively model the additional, e.g.,  radiative, transient, and diffusive processes present in these systems.

Thus, to extract information from the observed spectroscopic data, the data must be compared against model outputs to determine the conditions that generated the observations---i.e., the data processing takes the form of an inverse problem.  Because both the observation and the models describing the physical process that generate it are complicated and include uncertainties that are difficult to quantify---due to noise and observational errors but also modeling approximations and assumptions---it is challenging to represent their effects upon the resulting inferred quantities of interest.

While it is well within the scope and capabilities of statistical and computational science to provide solutions to this problem, the complexity and nuance of the problem itself alongside the significant laboratory challenge of simply getting experiments to function properly, can lead to a type of extreme instrumentalism \cite{leplin2017realism} in laboratory environments where the spectra and models are, e.g., treated in potentially perilously reductive ways, such as by hasty and subjective visual inspection performed `over-the-bench' by machine operators, to performing rudimentary (and potentially overly-simplistic) deterministic regression techniques, such as least squares residual analysis between the spectra and the kinetic model outputs.  These interpretive conditions are then often reported with weakly justified generic absolute error bars in the literature, leading to a concern that the reported results are not rigorously and reliably understood by the operators themselves, or, for that matter, sufficiently aligned with reproducible scientific modes of inquiry and compliance.

In this work, we focus on solution methods to this problem that utilize Bayesian analysis \cite{von2011bayesian} to provide principled and rigorous statistical methodologies, with the  motivating factors being to develop and analyze a framework that allows for the automation of uncertainty quantification in this specific context.  A broad overview of how Bayesian spectral fitting methods have been developed and implemented for spectroscopic data analysis can be found in \cite{cao2019bayesian,razul2003bayesian}, and more specifically the utilization of these types of models for evaluating emission spectra in plasma discharges in \cite{dodt2006analysis} and fusion devices \cite{fischer2010integrated,bozhenkov2017thomson,verdoolaege2006integrated,sciortino2020inference}.  Similar approaches can be found for gamma-ray spectroscopy  \cite{burr2020application}, Nuclear Magnetic Resonance spectroscopy \cite{rubtsov2007time,matviychuk2017experimental}, transmission spectroscopy \cite{pluriel2022toward,zingales2022toward}, diffuse reflectance spectroscopy \cite{hohmann2021analysis}, and quantum noise spectroscopy \cite{ferrie2018bayesian}.  Finally, mixture models are considered in \cite{park2009separating}, while automation techniques for these types of techniques are addressed in \cite{li2008neural}.

The specific motivating experiment for this work is the Plasma Liner Experiment (PLX) at Los Alamos National Laboratory, and particularly, the collection of emission spectra from the formed spherical plasma liners. Emission spectroscopy is a powerful diagnostic technique for the PLX liner plasmas, which are comprised largely of mid-density $10^{15}-10^{17}$ cm$^{-3}$ low temperature $1-5$~eV plasmas with typically argon species. The plasma emits a rich series of lines in the visible spectrum, which are readily collected via fiber-coupled spectrometers and digitized on CCD detectors, accumulating a large volume of data in the form of images.  Comparison of the emission lines to atomic physics models contains useful information on the liner electron temperature and ionization state evolution, but to date this comparison has been performed manually, by generation of a large number of comparison plots between observation and many candidate model spectra, and manual comparison to determine estimates of the plasma parameters.  This procedure has limited the amount of information that can be gleaned from the spectra and left them somewhat under-utilized on many experimental campaigns, motivating the development of a more powerful and automated method.

The structure of this paper is organized as follows.  Section~\ref{sec:sec_inference} describes the inference approach developed by integrating Bayesian inference techniques with base physical models. Sections~\ref{sec:sec_experimental} discusses the details of the experimental system under consideration, and Sections~\ref{sec:sec_results} shows the results of the application of the Bayesian inference approach to the experimental data interpretation on this experimental system.  Finally, we provide a conclusion and future directions in Section~\ref{sec:conclusion}.

\section{Inference Approach}
\label{sec:sec_inference}
To determine the state of the plasma---including, potentially, the
temperature, density, and composition---a model relating these
variables to the observed spectral data is required.  The plasma
properties of interest are then inferred based on what parameter values lead to
agreement between the modeled and observed spectra.  This process---i.e.,
determining the value that minimizes the misfit between the experiment
and the model---naturally takes the form of an optimization problem,
but a straightforward optimization does not automatically account for
uncertainty, which is introduced both by uncertainties inherent in any
measurement process as well as by approximations inherent in any
modeling process.  To account for these uncertainties, a Bayesian
inference approach is adopted here.  In this approach, the solution of
the inverse problem is a probability distribution for the plasma
properties of interest.  This probability distribution characterizes
not only the most likely values of the parameters but also the
relative probability of other values, accounting for the uncertainties
in the measurements and models underlying the analysis.

\subsection{Bayesian Inference Overview}
To be concrete, let $\theta$ denote the collection of plasma properties we hope to
infer.  Further, let $y_{obs}$ denote the observations, and let
$\mathcal{M}$ denote a physical model that maps the parameters
$\theta$ to the observations $y_{obs}$.  Given such a model, the approach here
is to infer the values of the parameters by finding those values that
make the output from the model best match the observations.  If both the model
and the observations were perfect---i.e., free from errors and
uncertainties---then there would exist values of the parameters,
denoted $\theta^{\ast}$, such that
\begin{equation*}
y_{obs} = \mathcal{M}(\theta^{\ast}),
\end{equation*}
and we could discover such parameter values by solving the following
minimization problem:
\begin{equation*}
\theta^{\ast} = \arg \min_{\theta} \| y_{obs} - \mathcal{M}(\theta) \|,
\end{equation*}
where $\| y_{obs} - \mathcal{M}(\theta) \|$ denotes a metric function (usually a norm) measuring the
discrepancy between the observations and the model.

Of course, in practice, it is almost never the case that both the
observations $y_{obs}$ and model $\mathcal{M}$ are entirely free from
errors and uncertainties.  In this situation, if only the ``best''
parameter values are of interest, then the optimization approach is
still feasible.  However, it is often of interest to find not only the
best fit but all values that lead to adequate, in some sense,
agreement between the measurements and the model and to characterize
the relative plausibility of these values.  When such uncertainties
are of interest, the Bayesian statistical framework provides a
coherent and self-consistent methodology for such analysis.

Bayesian methods have become common in a wide range of scientific
applications involving inference problems~\cite{Christian2001, jaynes2003probability, Kaipo2005, Calvetti2007, von2011bayesian, cheung2011bayesian,oliver2015validating}.  In the Bayesian framework,
the solution of the inference problem is given by the probability
density for the parameters of interest $\theta$ conditioned on the
observations $y_{obs}$---an object known as the ``posterior''
probability density.  Denoting this posterior by $\pi(\theta | y)$, Bayes' theorem
states that
\begin{equation*}
\pi(\theta | y) = \frac{\pi(\theta) \, \pi(y | \theta)}{ \pi(y) },
\end{equation*}
where $\pi(\theta)$ is known as the ``prior'' and $\pi(y | \theta)$ is
the ``likelihood''.  The denominator $\pi(y)$ is a normalizing
constant that ensures that the posterior integrates to one and will
not be discussed further here.  The prior characterizes any
information about the parameters that is available independent from
the observations $y_{obs}$.  For this work, the prior is taken to be a
uniform distribution over a set of possible plasma conditions selected
by the analyst.
To complete the formulation, the likelihood must be
specified.  The likelihood is based on two models: 1) the physical model
$\mathcal{M}$, which is described in section~\S\ref{sec:physical-model} and 2) a
probabilistic model representing all relevant uncertainties, which is
described in section~\S\ref{sec:probabilistic-model}.

\subsection{Physical Models} \label{sec:physical-model}
For the sake of exposition, the physical model under consideration here is split into two components: 1) the \emph{base physics model},
which determines the expected spectrum under given conditions, not accounting for measurement
imperfections, and 2) the \emph{measurement model}, which
determines the expected observed spectrum under given conditions and
outputs of the base physics model, and does account for potential
measurement distortions.

\subsubsection{Base physics model} \label{sec:base-physics-model}
Detailed modeling of emission spectra from plasmas is a broad and deep subject, and several modeling paradigms can be (and have been) applied depending on the plasma conditions of interest.  One paradigmatic categorization is whether the plasma can be described by certain equilibria, such as local-thermal-equilibrium (LTE) conditions \cite{fujimoto1990validity}, which hold in dense plasmas when the plasma collisions are frequent and fully govern electron energy state populations, and coronal-equilibrium conditions \cite{fujimoto1973validity}, which hold in rarefied plasmas where the dominant de-excitation of electrons is radiative relaxation to the ground state.  In contrast, at intermediate densities, the modeling paradigm of interest generally used to model the relative contributions of these mechanisms is a collisional-radiative (CR) model, which solves for the populations of electron energy states at given conditions and thus the relative strength of emission lines \cite{ralchenko2016modern}.

The simplest version of these models describe local (zero dimensional) light emission from a point source at a given plasma condition, but additional physical effects may occur en route to the observational instrument, such as those caused by optical opacity of the plasma, etc. These effects can be treated by techniques common in radiation transport, including, for example, radiation diffusion and characteristics methods, which can be solved simultaneously to determine the electron energy level populations. For the plasma conditions of interest in this paper, the plasma density is low enough that it can be assumed to be optically thin for the visible emission lines of interest, and effects of radiation transport can subsequently be neglected.

A complication for the analysis of spectra of optically thin plasmas to experimental measurements is that most practical experimental light collection schemes lead to the collection of light from a finite volume of the plasma, typically a narrow line-of-sight captured by a viewing lens or collimated-fiber-optic that delivers light from the plasma to a spectrometer. Thus, the recorded spectrum contains contributions from different spatial regions of the plasma, which may in general correspond to different physical (local) densities, temperatures, etc. As a consequence, a simplifying modeling assumption is made along the spatial profile of the discrete plasma conditions that assumes QoI homogeneity along the spectrometer's line of sight, thus allowing for the inverse problem to be solved over the entire line-integrated volume of plasma observed.  While it is possible to enrich this simplifying assumption by either modifying the underlying model (and hence usually its modeling assumptions) or by adding additional experimental diagnostics, the resulting problem is still more than likely to remain largely under-determined \cite{craig1976fundamental}.

In this paper the atomic spectra are obtained using the CR model PrismSPECT \cite{macfarlane2003simulation}.  PrismSPECT is a non-local (or local) thermodynamic equilibrium collisional-radiative model that calculates emission and absorption spectrum assuming a uniform plasma given a fixed electron temperature $T_e$ and density $\rho$ \cite{sawada2008experimental,han2015modeling}.  The Bayesian inference model utilized by SpectAI is independent of the underlying model generating the spectra that are being compared.  Thus, in principle, these inputs can be obtained using any atomic emission and absorption spectra model.  For example, other collisional radiative based models that could, in principle, be used include FLYCHK \cite{chung2003flychk,chung2005flychk}, ATOMIC \cite{hakel2006new}, ABAKO \cite{florido2009modeling}, OPAL \cite{iglesias1991opacities}, OPAS \cite{mondet2015opacity}, SCO-RCG \cite{nagayama2016model}, etc., or see \cite{piron2017review} for a broad overview and comparison of available codes/models.

\subsubsection{Measurement model} \label{sec:measurement-model}
The \emph{measurement model} utilizes the \emph{base physics model} while accounting for two additional separate issues.  First, the
instrument model introduces spectral broadening present beyond
the physical broadening mechanisms represented in the base physics model,
the instrument being limited in spectral resolution for example depending on 
the groove density of its diffraction grating.  Second, the measurement model accounts for the data conversion between what the
spectrometer records intensity in (i.e. as a function of pixel index), and the corresponding wavelength used for interpreting the corresponding base physical model output.  In principle, other effects could be treated, but we restrict to these two in the present work.

These measurement features are essentially characteristics of the
observation equipment---i.e., the spectrometer and any associated
infrastructure.  Thus, in principle, they can be determined by an
instrument calibration process to a known calibration source, which is
commonly done.  With a good calibration, one can then assume that
these features are simply fixed for the Bayesian analysis described
in \S\ref{sec:probabilistic-model}, which we do here.  However, since
even small mismatches between the observed and modeled spectra can
significantly affect the inference results, mismatches that could
occur for example from a slight rotation of the diffraction grating in
the positioning system, we have found it beneficial to include a
method to estimate these features using the observations of interest,
rather than using separate calibration data. These techniques are
described below.

\paragraph{Instrument line broadening}
The effect of instrument broadening is represented by convolving a
Gaussian function with the output of the model
from section~\S\ref{sec:base-physics-model}.  Specifically, given the output
of the base physics model, $\mathcal{M}_b(\lambda; \theta)$, the model
predicts that the spectrum that should be observed when accounting for
instrument broadening is given by
\begin{equation}
\mathcal{M}(\lambda; \theta, \sigma)
=
\int_{-\infty}^{\infty} \mathcal{M}_b(\tilde{\lambda}; \theta) \,
                        g(\lambda - \tilde{\lambda}; \sigma) \, \mathrm{d} \tilde{\lambda},
\end{equation}
where
\begin{equation}
g(s; \sigma) = \frac{1}{\sqrt{2 \pi \sigma^2}} \exp \left( -\, \frac{1}{2} \frac{s^2}{\sigma^2} \right),
\end{equation}
and the parameter $\sigma$ is the standard deviation, which determines
the broadening width.  

\paragraph{Pixel to wavelength mapping}
The intensity measurement is recorded as a function of pixel index.
In order to compare this to the model $\mathcal{M}(\lambda; \theta)$,
the pixel index $p$ must be mapped to the corresponding wavelength.
This is provided by a generic function $\lambda(p)$,
where the function $\lambda(p)$ is estimated from measurements recorded for a
known source.  Thus, this map is typically provided independent of the
measurements of interest.  In this work, we use a linear mapping:
\begin{equation}
\lambda(p) = m p + b,
\end{equation}
where the parameters $m$ and $b$ may be set by the analyst.  However,
it has been observed that even small deviations in the parameters $m$
and $b$, or small modeling errors that lead to shifts in the modeled
spectrum, can contaminate the inference process and lead to
temperature inferences that are implausible.  To avoid this problem,
the parameters $m$ and $b$ can also be calibrated using the observed
spectrum and the modeled spectrum at a selected condition.  As with
$\sigma$, this calibration could be incorporated into the Bayesian
inference, but this complication is beyond the scope of the present
work.  Instead, the parameters $m$ and $b$ are determined using a
deterministic least-squares solve using peak locations from a chosen
experiment and single simulation.  That is, given the highest $N$
intensity peaks at locations $\lambda_k$, $k = 1, \ldots, N$ in the
selected simulation and the corresponding $N$ highest peaks at pixels $p_k$, $k
= 1, \ldots, N$ in the observed spectrum, $m$ and $b$ are given
by
\begin{equation}
(m, b) = \arg \min_{\hat{m},\hat{b}} \sum_{k=1}^{N} \left( \lambda_k - \hat{m} p_k - \hat{b} \right)^2,
\label{eqn:ptow_least_squares}
\end{equation}
where the peaks are ordered according to increasing wavelength in the
simulation and increasing pixel in the observations.  This aligns the peaks in the 
selected simulation with those in the experiment as well as possible given the linear fit. 
Example results for this inference are shown in Fig.~\ref{fig:ptow_map}

\subsection{Probabilistic Model} \label{sec:probabilistic-model}
The resulting likelihood from the Bayesian procedure encodes the plausibility of the observation for given
values of the parameters, accounting for uncertainties in both the
observations and the model.  Thus, to formulate a likelihood, we
require a probabilistic model of the discrepancy between the model and
the observation.  Such a model could be the source of nearly endless
development and debate.  Here, we adopt a practical approach that does
not require detailed information about sources of error in either the
observation or the physical model.  As such, it is applicable in
situations where the details of the precise sources of error are not
well-understood, as is common in practice with complex experiments and
models.  However, the approach does account for known characteristics
of data produced by emission spectroscopy.  Specifically, we assume
that both the observations and the model predictions consist of a set
of strong ``lines'' embedded in a background signal.  An example of
such an observation is shown in Figure~\ref{fig:7330-spectrum}.  The
likelihood model used here is based upon a simple algebraic model of
functions that have such features.

The algebraic spectrum model takes the following form:
\begin{equation}
\mathcal{I}(\lambda) = \sum_{i=1}^{N} s_i g(\lambda; \lambda_i, w_i) + b(\lambda),
\label{eqn:spectrum_model}
\end{equation}
where $\mathcal{I}$ is the spectral intensity, $\lambda$ is the
wavelength, $N$ is the number of lines, $s_i$ is the strength of the
$i$th line, $g$ is a line shape function, $\lambda_i$ is the location
of the $i$th line, $w_i$ is the width of the $i$th line, and $b$ is
the background.  For the remainder of this work, we assume that the
background is negligible or can be exactly removed prior to the
analysis, and thus it will be neglected going forward.

In designing a likelihood model, the task is to
assign probabilities to observed differences.  Toward this end,
consider the difference between two spectra.  Using superscripts to
distinguish between these two spectra and
assuming~\eqref{eqn:spectrum_model} is appropriate for both, their
difference can be written as
\begin{equation*}
\delta \mathcal{I}(\lambda)
=
\mathcal{I}^{(1)}(\lambda) - \mathcal{I}^{(0)}(\lambda)
=
\sum_{i=1}^{N_1} s^{(1)}_i g^{(1)}(\lambda; \lambda^{(1)}_i, w^{(1)}_i)
-
\sum_{i=1}^{N_0} s^{(0)}_i g^{(0)}(\lambda; \lambda^{(0)}_i, w^{(0)}_i).
\end{equation*}
Spectra that are ``close'' will have the same number of lines, $N_1 =
N_0 = N$, and be described using the same line shape functions
$g^{(1)} = g^{(0)} = g$.  In this case,
\begin{equation*}
\delta \mathcal{I}(\lambda)
=
\sum_{i=1}^{N}
s^{(1)}_i g(\lambda; \lambda^{(1)}_i, w^{(1)}_i) -
s^{(0)}_i g(\lambda; \lambda^{(0)}_i, w^{(0)}_i)
\end{equation*}
Finally, we can further simplify by linearizing this result---i.e., by
writing spectrum (1) as a Taylor series expansion around spectrum
(0) parameters and dropping all terms higher than first order.  Specifically,
\begin{align*}
\delta \mathcal{I}(\lambda)
& =
\sum_{i=1}^{N}
(s^{(0)}_i + \delta s_i) (g(\lambda; \lambda^{(0)}_i + \delta \lambda_i, w^{(0)}_i + \delta w) -
s^{(0)}_i g(\lambda; \lambda^{(0)}_i, w^{(0)}_i) \\
& \approx
\sum_{i=1}^{N}
\delta s_i g(\lambda; \lambda_i, w_i)
+
s_i \frac{\partial g}{\partial \lambda_i} \delta \lambda_i
+
s_i \frac{\partial g}{\partial w_i} \delta w_i,
\end{align*}
where $\delta s_i = s_i^{(1)} - s_i^{(0)}$, $\delta \lambda_i
= \lambda_i^{(1)} - \lambda_i^{(0)}$, $\delta w_i = w_i^{(1)} -
w_i^{(0)}$.

This development serves as the basis for the probabilistic model of
the discrepancy between the observed and modeled spectra.
Specifically, the model that follows is given by:
\begin{equation}
\mathcal{I}_{obs}(\lambda) - \mathcal{M}(\lambda; \theta)
=
\sum_{i=1}^{N}
\delta s_i g(\lambda; \lambda_i, w_i)
+
s_i \frac{\partial g}{\partial \lambda_i} \delta \lambda_i
+
s_i \frac{\partial g}{\partial w_i} \delta w_i
+
\epsilon
\label{eqn:discrepancy-model}
\end{equation}
where
\begin{gather*}
\delta s_i \sim N(0, \eta^2),\quad
\delta \lambda_i \sim N(0, \ell^2), \quad
\delta w_i \sim N(0, \sigma^2),
\end{gather*}
and $\epsilon$ is a zero-mean white noise Gaussian process.  Clearly
this model has a number of parameters: $N$; $s_i$, $\lambda_i$, and
$w_i$ for $i = 1, \ldots, N$; $\eta$, $\ell$, $\sigma$, and $\beta$.
Determination of these hyperparameters is described below.  Once these
values are fixed, the probabilistic model is entirely defined by the
distributions for $\delta s_i$, $\delta \lambda_i$, $\delta w_i$
and $\delta b$, which are Gaussian, the
form~\eqref{eqn:discrepancy-model}, which is linear in these
parameters, and the additive Gaussian noise $\epsilon$.  Thus, the
resulting probabilistic model of the discrepancy and the resulting
likelihood is a Gaussian process (GP).

\subsection{Hyperparameters} \label{sec:hyperparameters}
The parameters of the likelihood model, i.e., the hyperparameters, can
be split into two qualitatively different groups.  The first group
includes $N$, $s_i$, $\lambda_i$, and $w_i$.  These parameters characterize the
original spectrum that the likelihood model is describing deviations
from.  As such, these parameters are identified based on the original
observation.  We use standard, ``off-the-shelf'' techniques to
determine these parameters.  Specifically, we use the
\verb+find_peaks+ function from the \verb+scipy.signal+ package \cite{virtanen2020scipy}, that utilizes standard peak prominence heuristics to determine peak locations.  To avoid spurious peaks due to noise in the observations, peaks are required to have a height that is at least 0.03 times the maximum intensity in the observation window, and peaks are required to have a prominence of 0.01 times the maximum intensity in the observation window.

The second group of parameters includes $\eta$, $\ell$, $\sigma$, and
$\beta$.  These parameters control how the stochastic model assigns
probability around the result of the deterministic model.  These
parameters are determined in a preprocessing step that occurs prior to
the Bayesian inference.  First, reasonable values of the physical
model parameters are estimated using standard least-squares
minimization.  Then, denoting these initial parameter values by $\theta^*$,
the misfit $\varepsilon^*$ is computed via $\varepsilon^* = \mathcal{I}_{obs}
- \mathcal{M}(\theta^*)$. Given this misfit, the parameters $\delta
s_i$, $\delta \lambda_i$, $\delta w_i$, and $\delta b$ can be
determined by solving a linear least-squares problem to minimize the
discrepancy between the misfit model~\eqref{eqn:discrepancy-model} and the computed
$\varepsilon^*$.  Specifically, we solve the following optimization problem
\begin{equation}
\delta s^*_i, \delta \lambda^*_i, \delta w^*_i
=
\arg \min_{\delta s_i, \delta \lambda_i, \delta w_i} \| \varepsilon^* - e(\delta s_i, \delta \lambda_i, \delta w_i) \|,
\end{equation}
where
\begin{equation}
e(\delta s_i, \delta \lambda_i, \delta w_i)
= \sum_{i=1}^{N}
\delta s_i g(\lambda; \lambda_i, w_i)
+
s_i \frac{\partial g}{\partial \lambda_i} \delta \lambda_i
+
s_i \frac{\partial g}{\partial w_i} \delta w_i.
\end{equation}
This leads to $N$ (the number of lines) values for each parameter.
Then, the standard deviation paramters $\eta$, $\ell$, and $\sigma$
are determined by taking the observed standard deviation of this best
fit over all the lines.


\subsection{Summary} \label{sec:approach-summary}
To summarize the overall approach, the solution of the inverse problem
is given by the posterior probability distribution for the plasma
density and temperature, $\rho$ and $T$, given the observed intensity,
$I$:
\begin{equation*}
\pi(\rho, T | I) \propto \pi(\rho, T) \, \pi(I | \rho, T),
\end{equation*}
where the prior $\pi(\rho, T)$ is uniform over a range of conditions
and the likelihood $\pi(I | \rho, T)$ which incorporates the physical, measurement, and probabilistic models described in
\S~\ref{sec:physical-model}-\S\ref{sec:hyperparameters}.  Since the prior is taken to be
uniform here, the posterior is proportional to the likelihood
function.  The likelihood model is Gaussian in the intensity, but,
because the forward model relating the density and temperature to the
intensity is nonlinear, the resulting posterior probability over
density and temperature is not necessarily Gaussian and cannot be
sampled easily.  While specialized algorithms are available for
exploring such distributions, such as various Markov chain Monte Carlo
(MCMC) methods~\cite{martino2018review}, these algorithms require the evaluation of the
forward model inside the sampling loop.  Since the forward model
evaluation is the most computationally expensive component of this
approach, and since the space of interest is only two-dimensional, we
choose a grid-based approach instead.  In this approach, the model is
run on a grid in density, temperature space \textit{a priori} and then
the posterior is evaluated on that same grid.  In this way, a single
set of simulations can be used to support the inference for a large
number of experimental observations, which effectively amortizes the
computational effort over the entire experimental campaign and
dramatically reduces the time required for the inference on any single
experiment.  Thus, given an observed spectrum, the steps to evaluate
this posterior probability are as follows:
\begin{enumerate}
\item Run a set of physical model simulations---here using PrismSpect,
  as described in Section~\ref{sec:physical-model}---to compute the
  predicted spectrum for a set of plausible $\rho_i, T_i$ conditions,
  assuming the gas composition is known.
\item Determine the measurement model hyperparameters (see Section~\ref{sec:measurement-model}), including:
  \begin{itemize}
  \item The line broadening parameter, $\sigma$, and
  \item The pixel to wavelength mapping parameters, $m$ and $b$.
  \end{itemize}
\item Determine the probabilistic model hyperparameters, as discussed in Section~\ref{sec:hyperparameters}.
\item Evaluate the posterior at points where the simulations where
  run, $\pi(\rho_i, T_i | I)$, by evaluating the likelihood given the
  simulation results and observations, as described in
  Section~\ref{sec:probabilistic-model}.
\end{enumerate}
As discussed previously, step 1 need only be performed once for a
given set of experiments at similar conditions.

\section{Experimental Observations}
\label{sec:sec_experimental}

The inference framework described above is applied to the analysis of results from a series of experiments at PLX.  In these experiments, two unmagnetized argon plasma jets are collided head-on at varied (relative fixed) velocity and density, to study the collisionality of the supersonic plasmas under these conditions.  The plasma behavior under these conditions is relevant to the prospects of PLX / PJMIF as a spherical compression driver, specifically as related to occurrence of collisional plasma shocks vs. kinetic interpenetration of the jets. These phenomena have been observed in previous studies on PLX. Emission spectroscopy is collected from these experiments, and is an important diagnostic in the overall experiment for constraining the plasma electron temperature and average ionization state $\bar Z$.

\begin{figure}
  \centering
  \includegraphics[width=.9\linewidth]{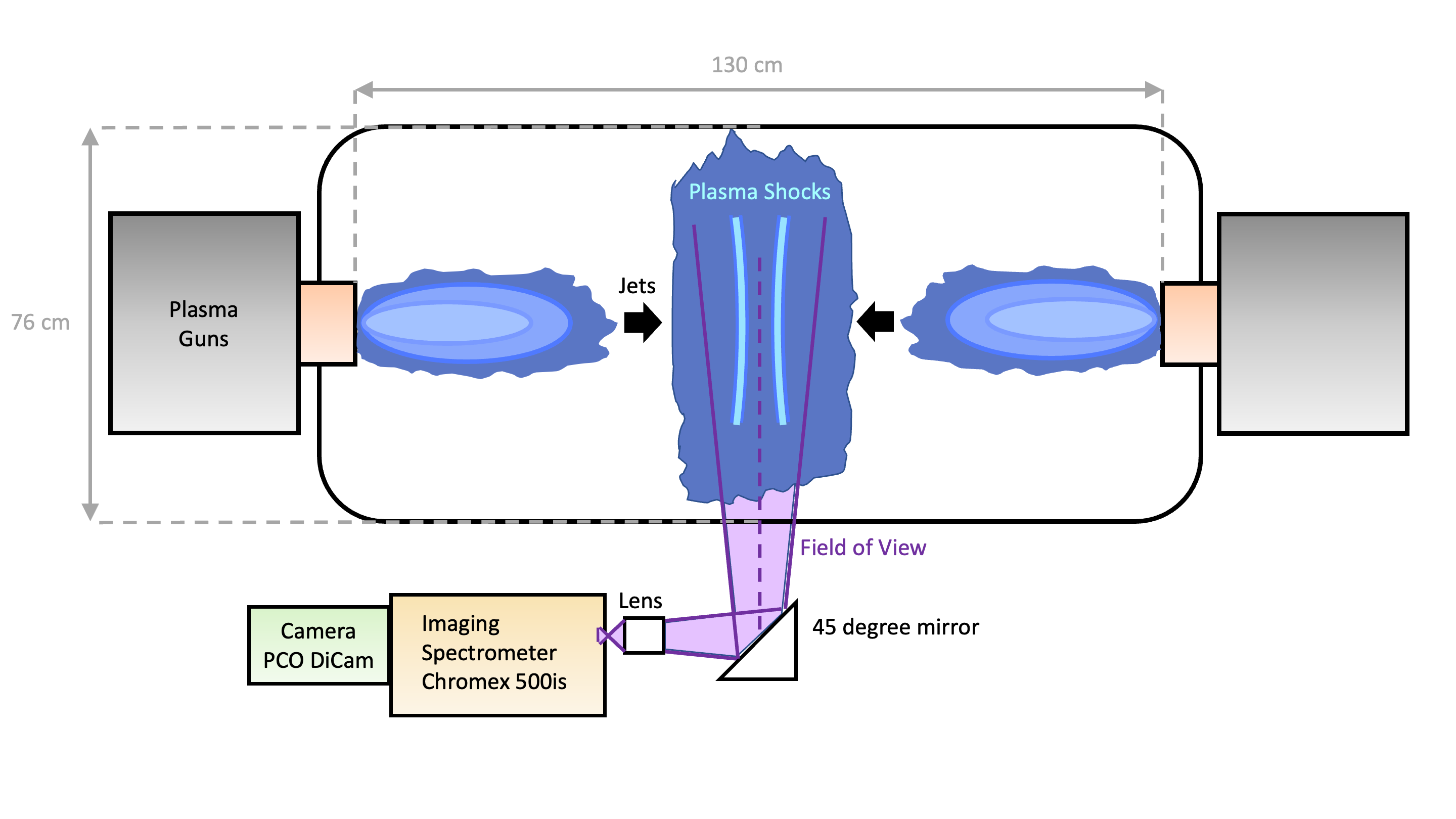}
  \caption{A diagram of the experimental setup of the jet merging chamber that uses plasma guns from PLX. The purple highlighted region indicates the spectrometer field of view.}
  \label{fig:experimentsetup}
\end{figure}

Figure~$\ref{fig:experimentsetup}$ shows a diagram of the experimental setup and spectroscopic diagnostic layout. Light is collected from the central volume of the colliding plasma jets and imaged onto the entrance slit of a Chromex 500is imaging spectrometer, which is coupled to a PCO Dicam Pro intensified CCD camera.  The spatial direction resolved by the spectrometer is aligned with the jet collision axis, and it was envisioned that this would provide spatially resolved data across plasma shocks when they occurred. In implementation, spatial resolution was significantly blurred by effects of perspective distortion in the entrance optics viewing the transparent plasma, so effective spatial resolution was not obtained. The data therefore constitute a spatially-integrated measure over the plasma volume.  Data are temporally resolved by control of the camera exposure time and delay, acquiring one frame of data per shot.  Shots are repeated with varied camera delays to build up a time-resolved evolution of the emission spectrum at each jet operating condition.

\begin{table}
\begin{center}
\begin{tabular}{ c|c|c|c }
Shot numbers & P inlet (psig) & V (kV) & avg. jet velocity (km/s) \\
 \hline
7328-7365 & 50 & 4 & $16 \pm 1.7$ \\
7366-7398, 7497-7504 & 30 & 4.5 & $28 \pm 4.1$ \\
7399-7433 & 40 & 4.25 & $26 \pm 5.6$ \\
7434-7470 & 45 & 4.25 & $23 \pm 3.7$ \\
7471-7496 & 35 & 4.25 & $27 \pm 5.9$ \\
7505-7540 & 20 & 4.5 & $52 \pm 1.5$\\
\end{tabular}
\end{center}
\caption{Experimental shot settings, and average measured plasma jet velocity}
\label{tab:exp_metadata}
\end{table}

A series of shots are performed spanning different jet operating conditions, summarized in Table~\ref{tab:exp_metadata}. The main experimental settings that are varied are the charge voltage of the plasma gun capacitor bank, and the inlet pressure to the gas puff valve.  With higher bank voltage, more accelerating current is delivered, resulting in a higher velocity.  With higher inlet pressure, more gas is injected into the gun, resulting in a heavier jet and ultimately a slower velocity.

\begin{figure}[htp]
\begin{center}
\includegraphics[width=0.4\linewidth]{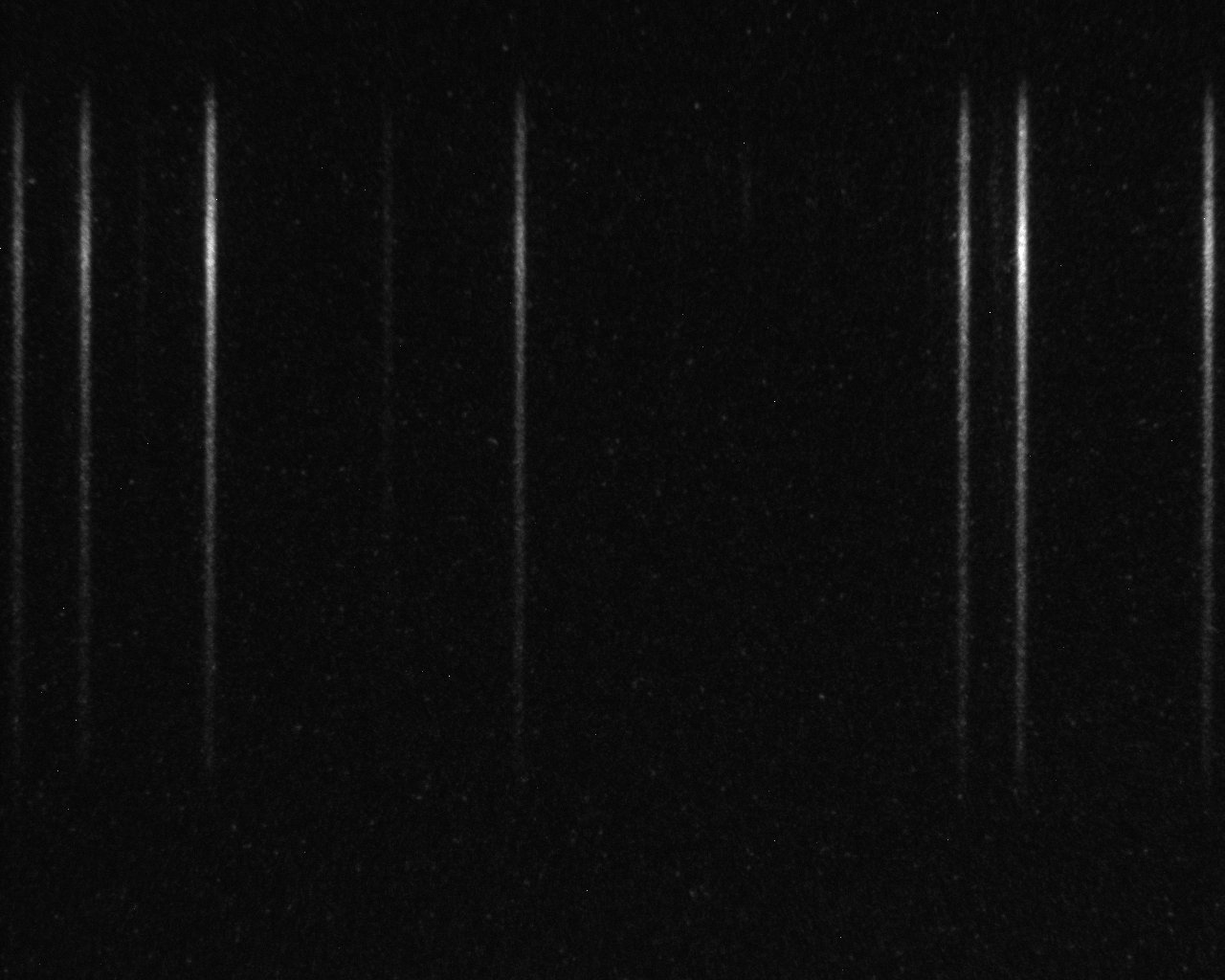}
\includegraphics[width=0.55\linewidth]{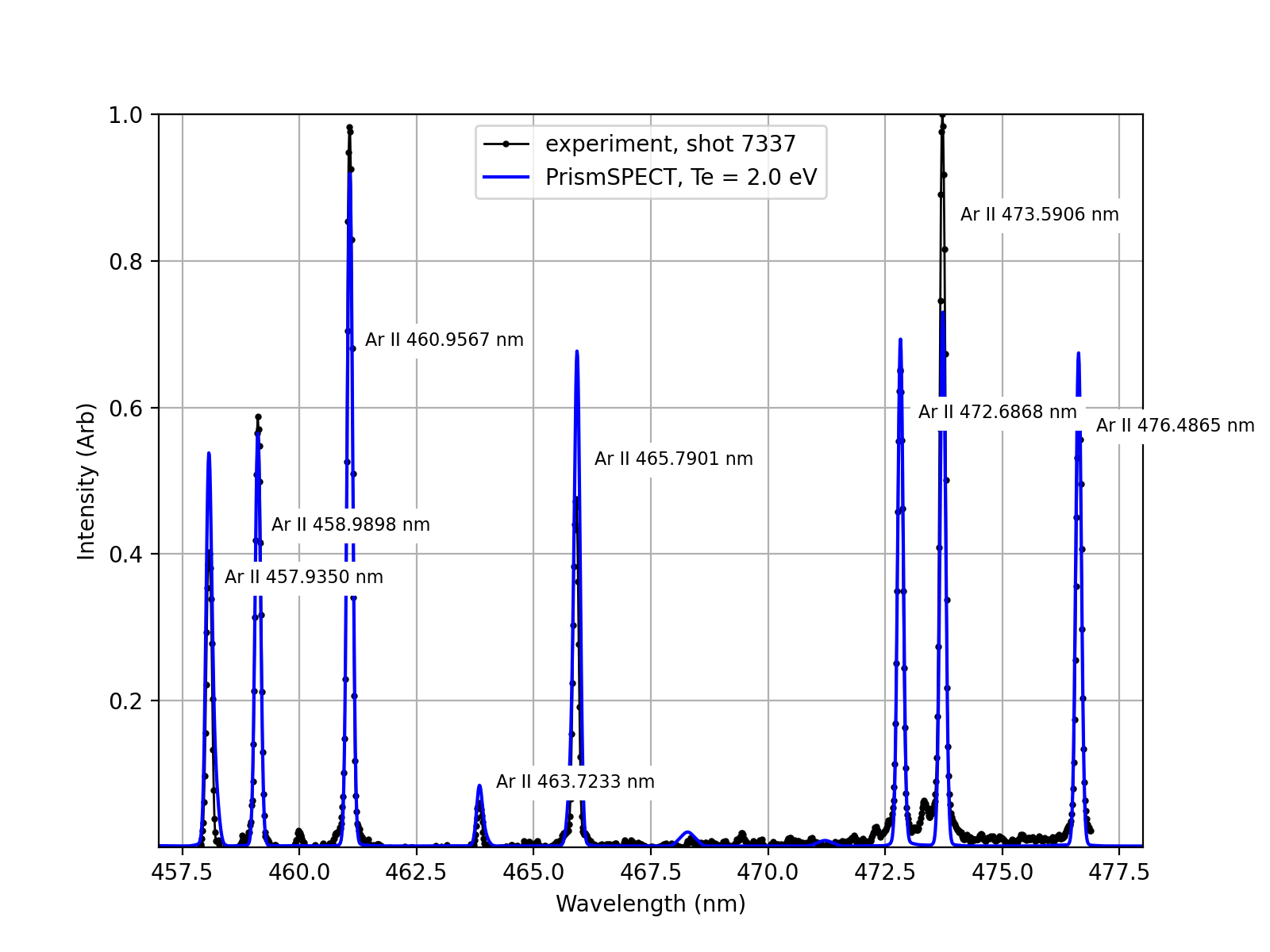}
\end{center}
\caption{(a) Raw data image from shot 7337.  (b) Processed and binned data from shot 7337, overlaid with PrismSPECT simulated spectrum. Visible strong lines of singly-ionized argon labeled.}
\label{fig:lineout_vs_pspect}
\end{figure}

Figure~$\ref{fig:lineout_vs_pspect}$ shows raw data obtained from an experimental shot, and the same data processed compared to a Prismspect simulated spectrum. Strong lines of singly-ionized argon are consistently observed across many experimental operating conditions, with relatively subtle changes in the spectrum from shot to shot. This is expected, as the radiative cooling from the ionized argon increases significantly with temperature in this range $\sim 2$ eV, and acts to clamp the temperature from rising greatly above this level.  A colder liner temperature is typically desireable in PLX / PJMIF to keep liner sound speed and density expansion low, and to increase the liner Mach number. This dataset is analyzed using the developed inference approach.

\section{Results}
\label{sec:sec_results}
The analysis can be decomposed into two phases: (1) a pre-processing
phase, where the line broadening and pixel to wavelength map
parameters (\S\ref{sec:physical-model}) and probabilistic model
hyperparameters (\S\ref{sec:hyperparameters}) are determined, and (2)
the Bayesian inference phase.  Sample results from each phase are
presented for cases identified in Table~\ref{tab:exp_metadata}.

\subsection{Pre-processing Phase Results}
\begin{figure}[htp]
\begin{center}
\begin{subfigure}[t]{0.49\linewidth}
  \centering
  \includegraphics[width=\linewidth]{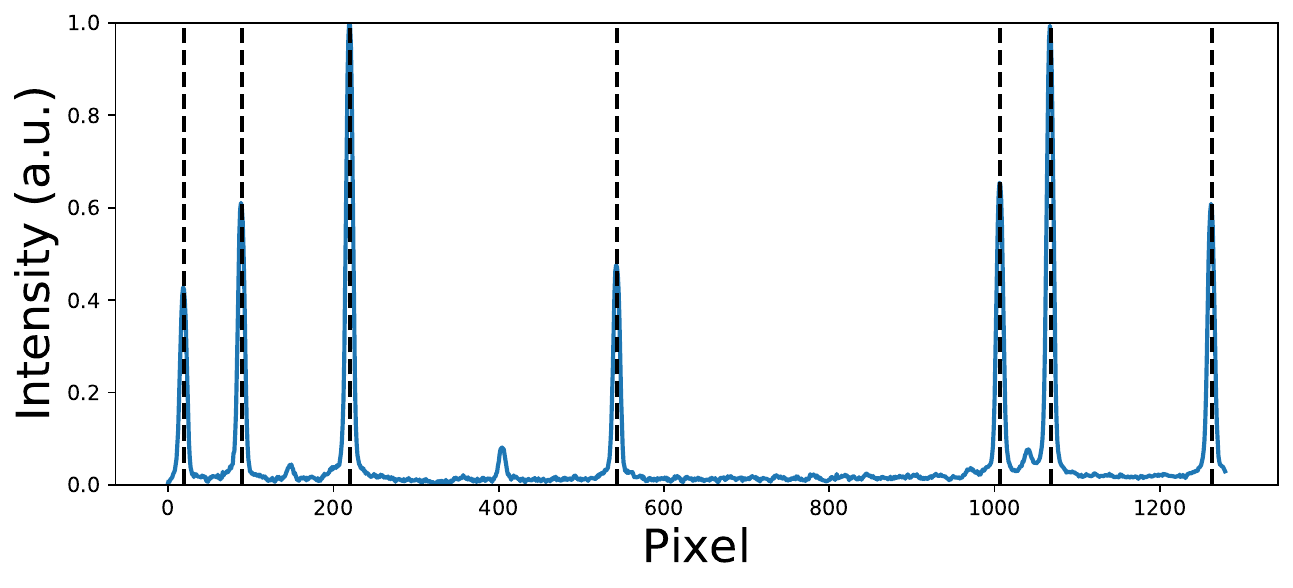}
  \caption{Observed intensity versus pixel for PLX shot 7330.}
\end{subfigure}
\hfill
\begin{subfigure}[t]{0.49\linewidth}
  \centering
  \includegraphics[width=\linewidth]{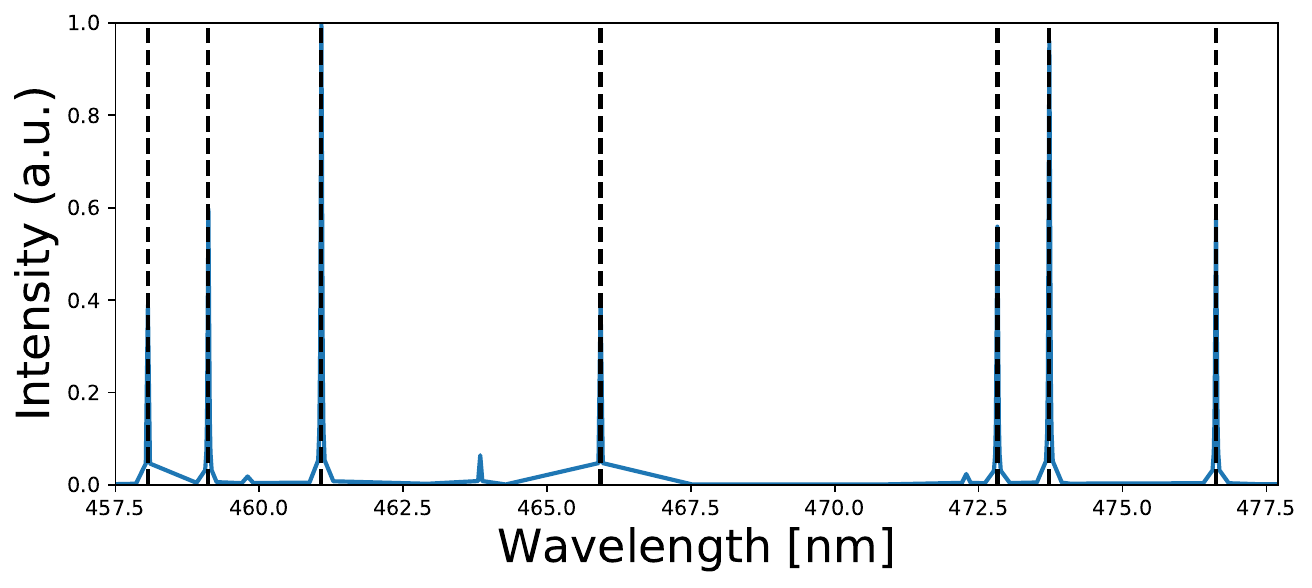}
  \caption{Modeled intensity versus wavelength example.}
\end{subfigure}
\end{center}
\caption{Intensity versus pixel and wavelength with peaks indicated by the vertical dashed lines.}
\label{fig:peak_locations}
\end{figure}

\begin{figure}[htp]
\begin{center}
\includegraphics[width=0.6\linewidth]{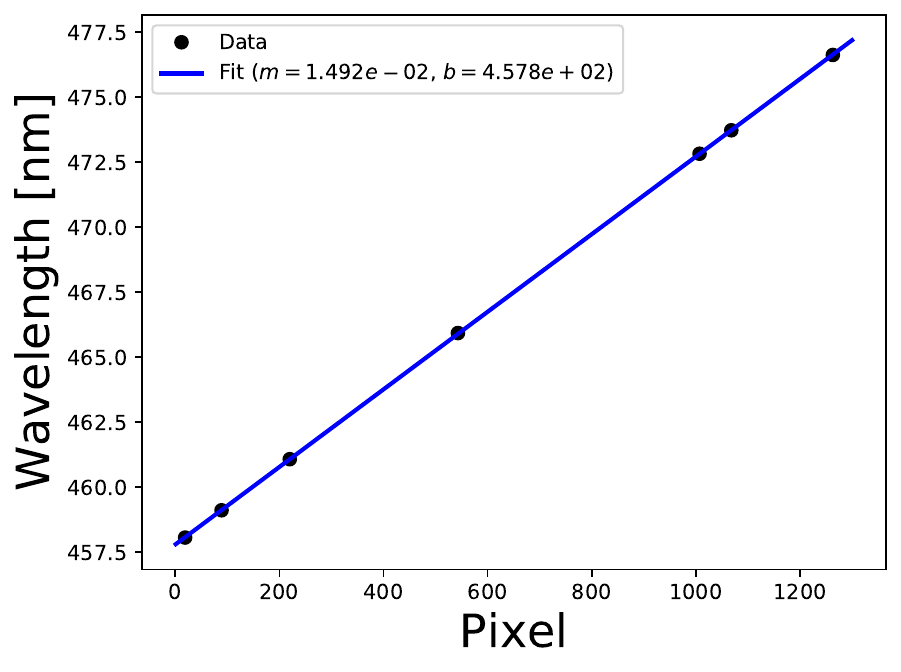}
\end{center}
\caption{Pixel to wavelength mapping determined by fitting peak wavelengths as a linear function of peak pixels.}
\label{fig:ptow_map}
\end{figure}

To determine the pixel to wavelength mapping, the least-squares
problem defined in~\eqref{eqn:ptow_least_squares} is solved.  The data
for this problem is determined using standard peak finding
techniques%
\footnote{Specifically, we use the \texttt{find\_peaks} function from
  the \texttt{scipy.signal} package.  See
  \url{https://docs.scipy.org/doc/scipy/reference/generated/scipy.signal.find\_peaks.html}}.%
~Figure~\ref{fig:peak_locations} shows example results from the peak finding phase for
solving the resulting least-squares problem gives the pixel with
wavelength mapping shown in Figure~\ref{fig:ptow_map}.  The
fit succeeds for this case and there is no suggestion that a
richer parameterization of the mapping is necessary.

\subsection{Bayesian Inference Results}
The inference results are split into two sections.  First, to give a
sample of the full picture of information that is obtained, joint
posteriors for density and temperature for individual shots are shown.
The features observed in the posterior as well as the fits between the
PrismSpect predictions and the measured spectra are examined.  Second,
we use the results to look for trends in the measured temperature as a
function of the experimental settings.  These results reinforce the
importance of accounting for uncertainty in the analysis of such
measurements.

\subsubsection{Individual shot results}
\begin{figure}[p]
\begin{center}
\begin{subfigure}[t]{0.45\linewidth}
  \centering
  \includegraphics[width=\linewidth]{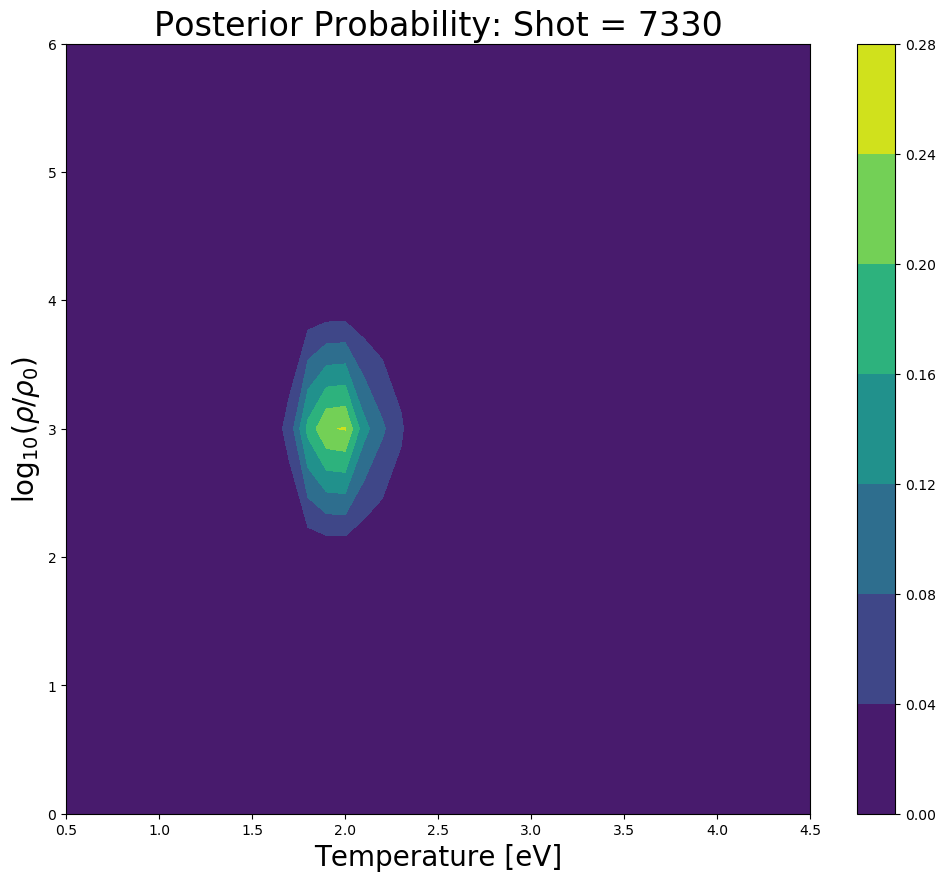}
\end{subfigure}
\hfill
\begin{subfigure}[t]{0.45\linewidth}
  \centering
  \includegraphics[width=\linewidth]{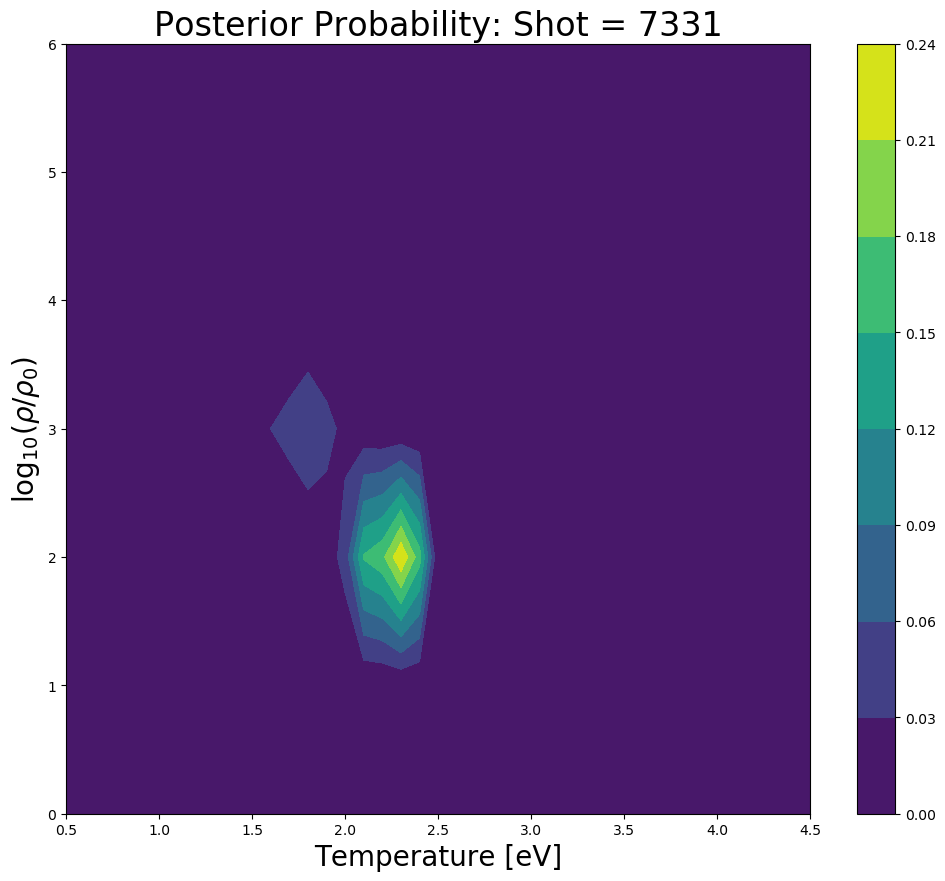}
\end{subfigure}
\begin{subfigure}[t]{0.45\linewidth}
  \centering
  \includegraphics[width=\linewidth]{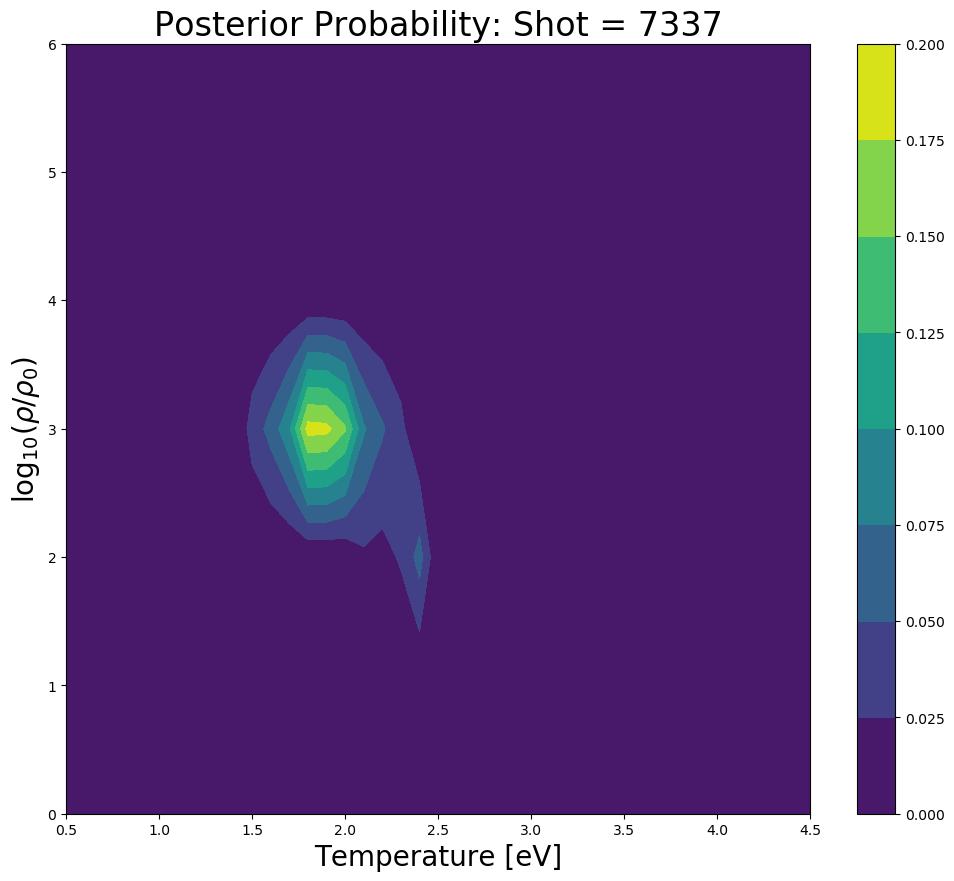}
\end{subfigure}
\hfill
\begin{subfigure}[t]{0.45\linewidth}
  \centering
  \includegraphics[width=\linewidth]{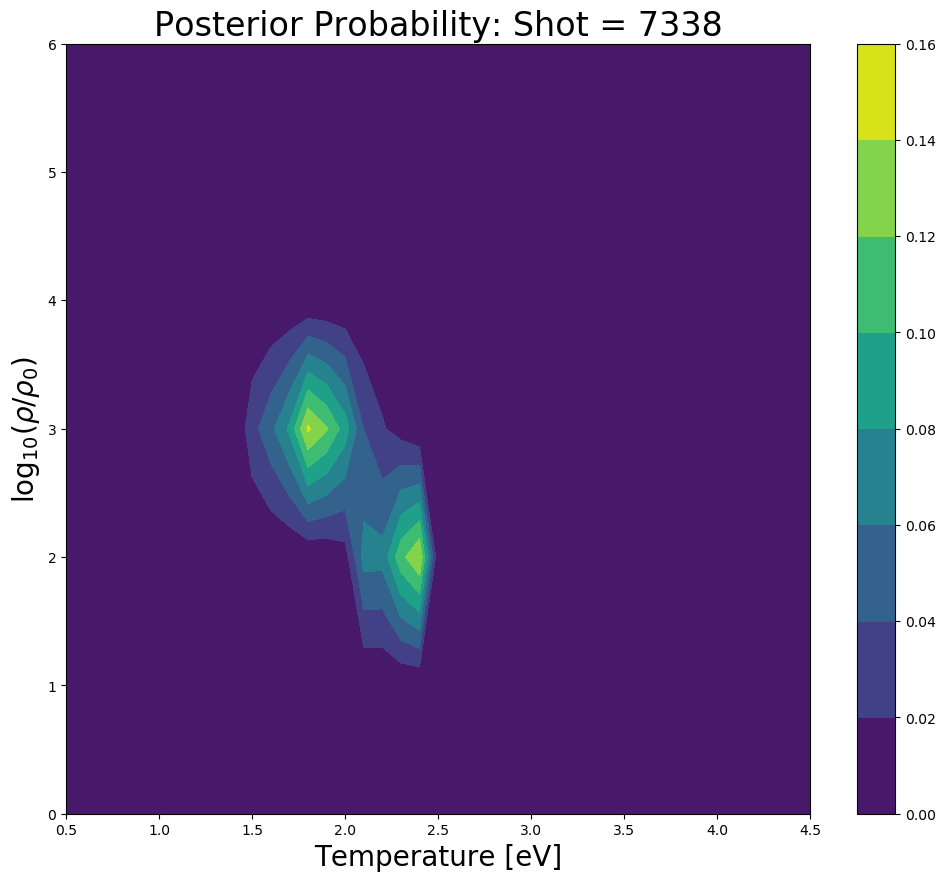}
\end{subfigure}
\begin{subfigure}[t]{0.45\linewidth}
  \centering
  \includegraphics[width=\linewidth]{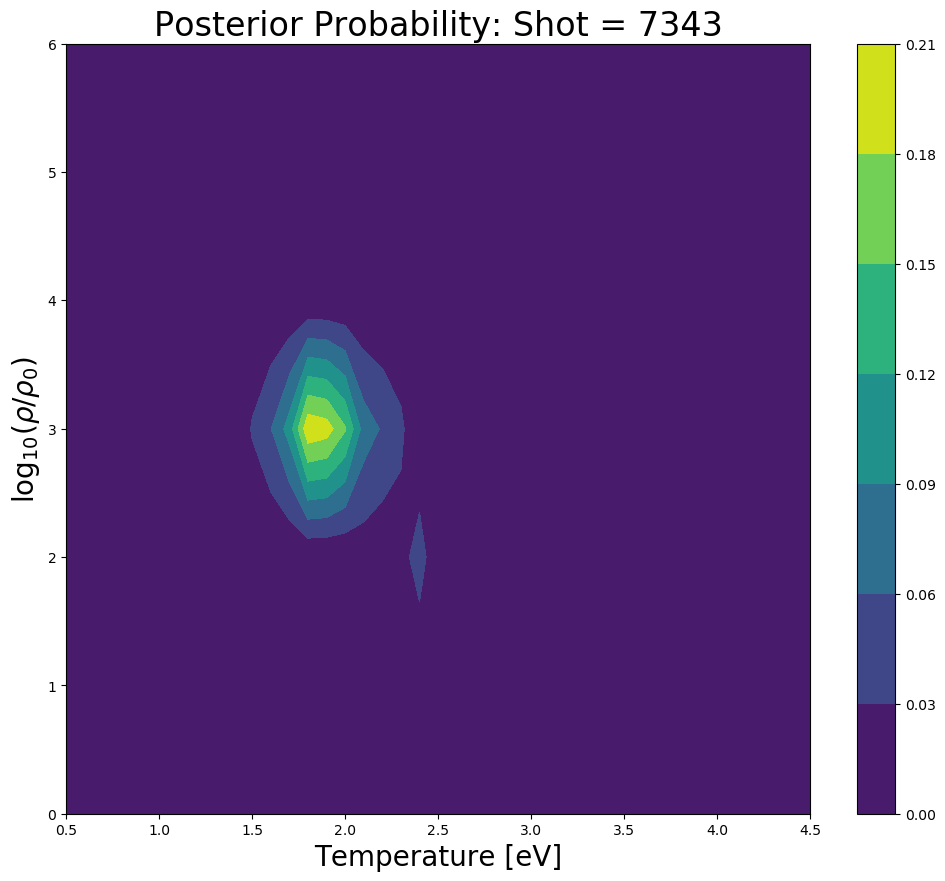}
\end{subfigure}
\hfill
\begin{subfigure}[t]{0.45\linewidth}
  \centering
  \includegraphics[width=\linewidth]{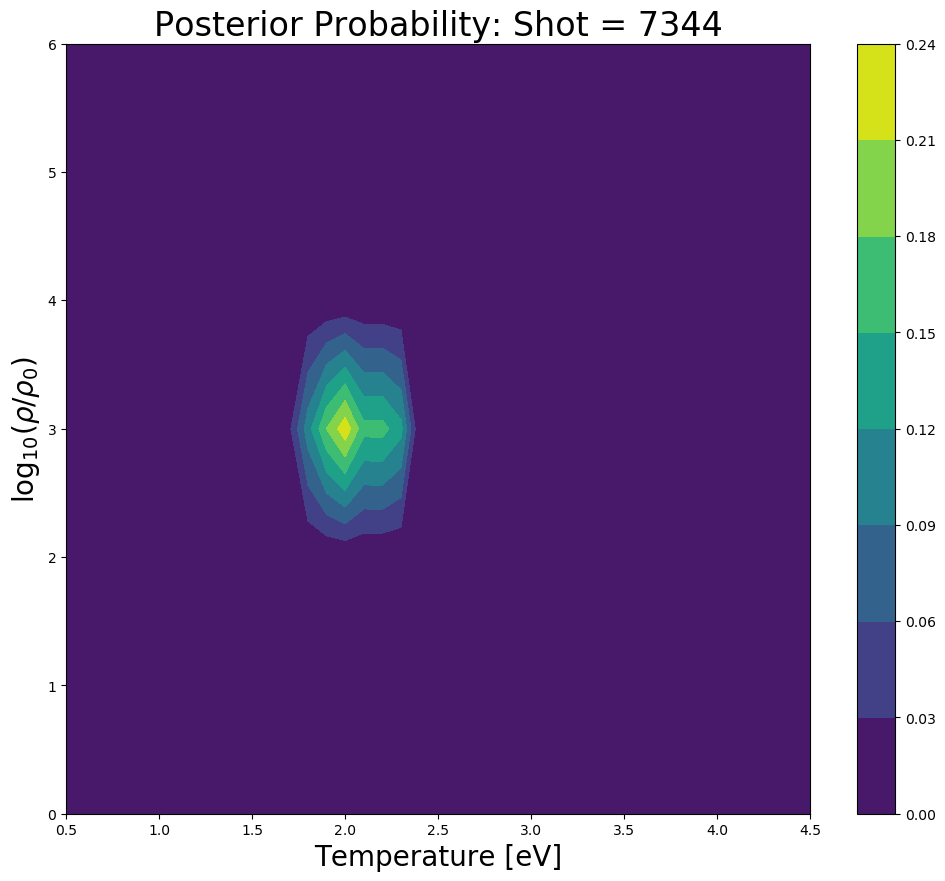}
\end{subfigure}
\end{center}
\caption{Joint posterior probability distributions for temperature and
  density inferred from spectra recorded during six PLX shots on
  10/1/2021.}
\label{fig:sample-joint-posteriors}
\end{figure}

Figure~\ref{fig:sample-joint-posteriors} shows the joint posterior
probability for six shots taken on October 1, 2021.  These
shots---specifically, 7330, 7331, 7337, 7338, 7343, and 7344---are all
at the same nominal conditions of 50 psig inlet pressure for the gas
puff valve and 4 kV voltage of the plasma gun capacitor bank.  The
spectra are measured at different camera delay times, with 7330 and
7331 using a delay of 58 $\mu$s while 7337 and 7338 have a 52 $\mu$s
delay and 7343 and 7344 have a 46 $\mu$s delay.  Dependencies on
camera delay time and other parameters are examined more closely in
Section~\ref{subsec:shots}. Here, these shots are used as representative
individual spectra, and the results
show some notable features.  First, the
uncertainty is not negligible.  The range of non-negligible posterior
probability generally spreads over an interval of approximately 0.5
eV.  Second, the posterior probability does not always concentrate
around a single peak.  The results from shots 7337 and 7338 show this
clearly.  Since the Bayesian approach provides a full probability
distribution as the solution of the inverse problem, such behavior is
allowed when multiple conditions fit the data similarly well, even if
they are separated in parameter space.

\begin{figure}[p]
\begin{center}
\begin{subfigure}[t]{0.45\linewidth}
  \centering
  \includegraphics[width=\linewidth]{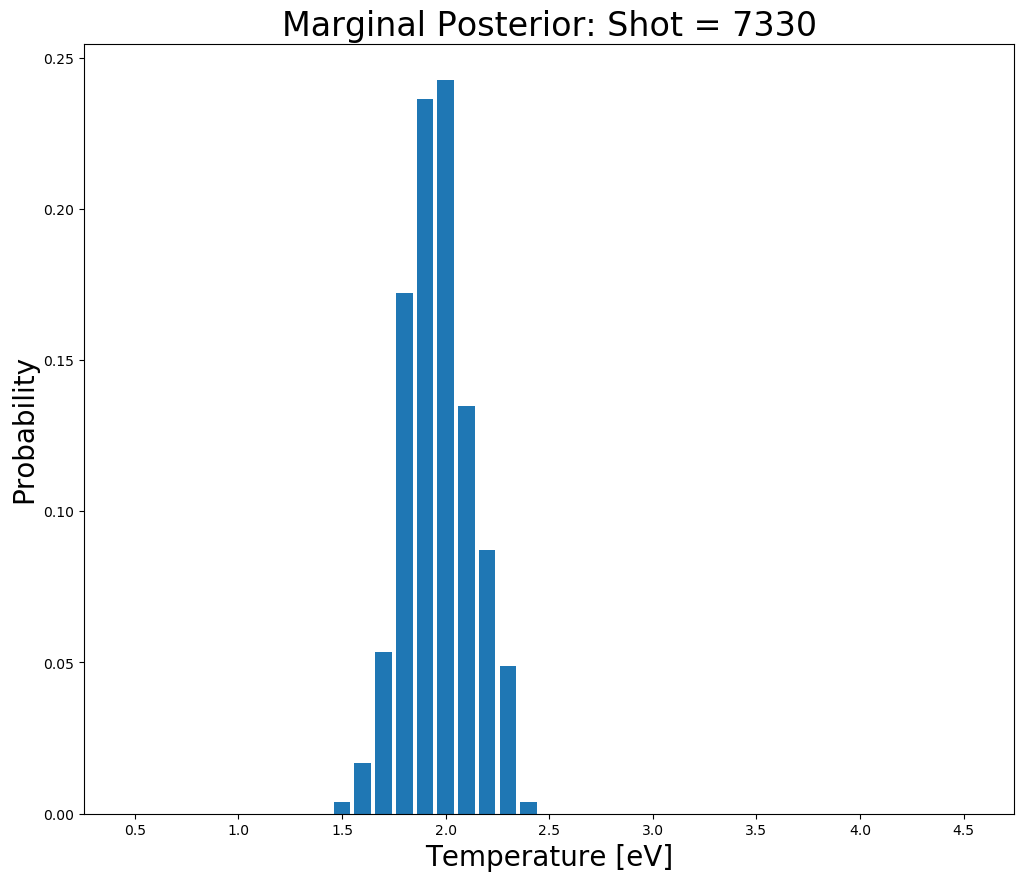}
\end{subfigure}
\hfill
\begin{subfigure}[t]{0.45\linewidth}
  \centering
  \includegraphics[width=\linewidth]{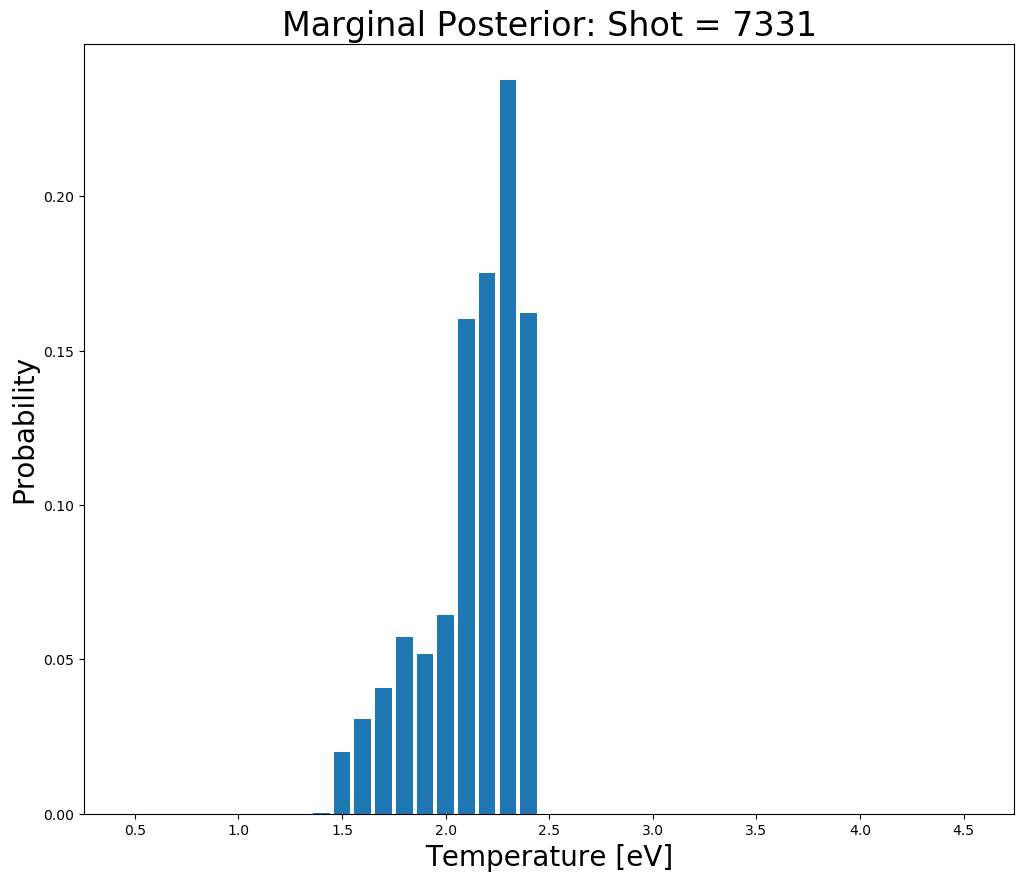}
\end{subfigure}
\begin{subfigure}[t]{0.45\linewidth}
  \centering
  \includegraphics[width=\linewidth]{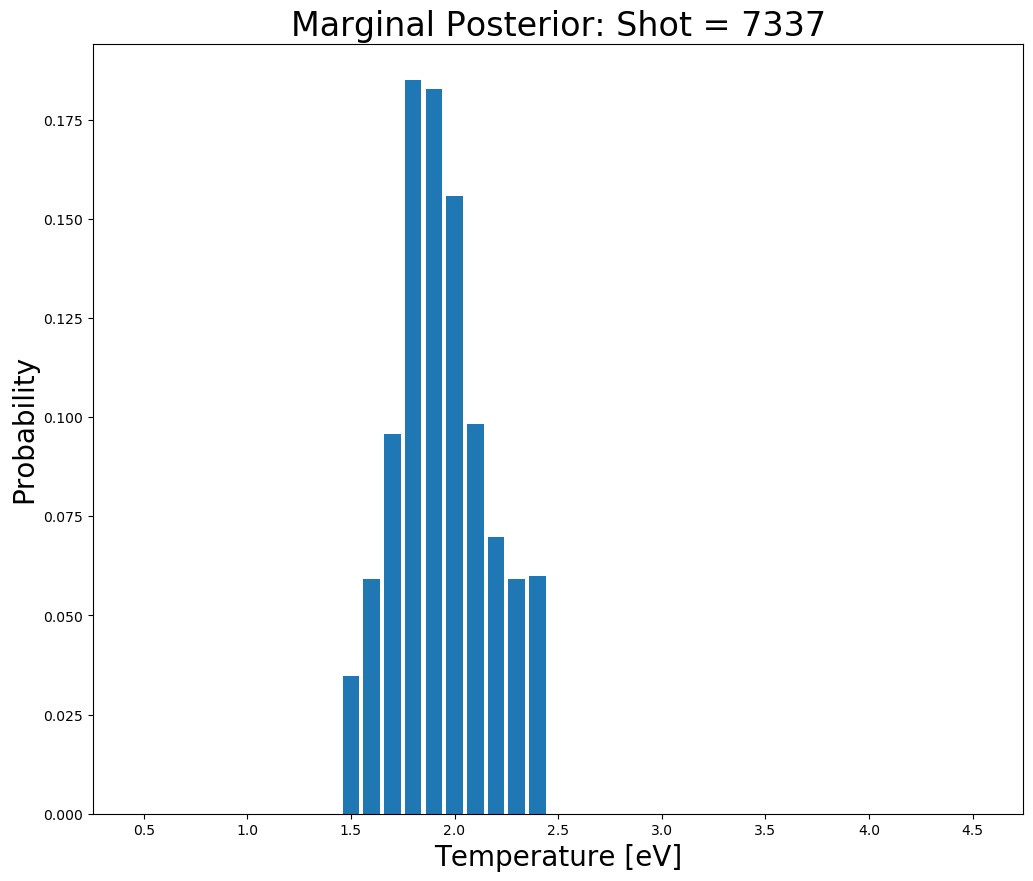}
\end{subfigure}
\hfill
\begin{subfigure}[t]{0.45\linewidth}
  \centering
  \includegraphics[width=\linewidth]{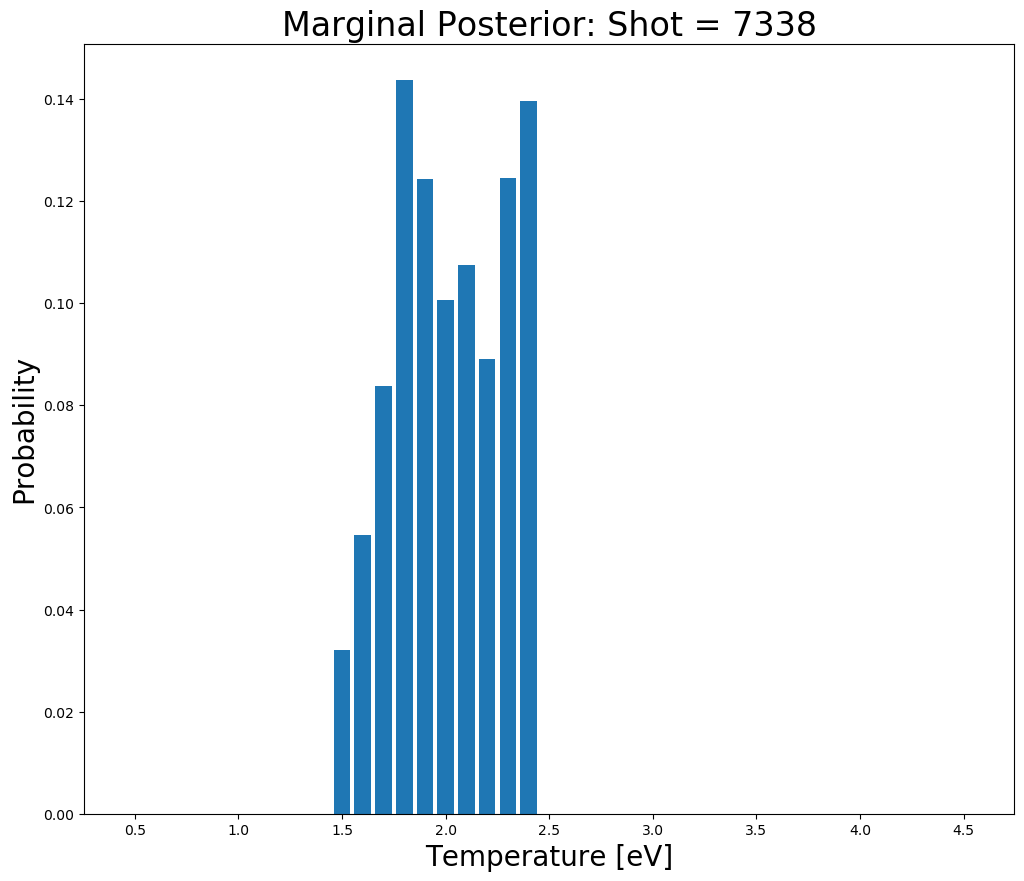}
\end{subfigure}
\begin{subfigure}[t]{0.45\linewidth}
  \centering
  \includegraphics[width=\linewidth]{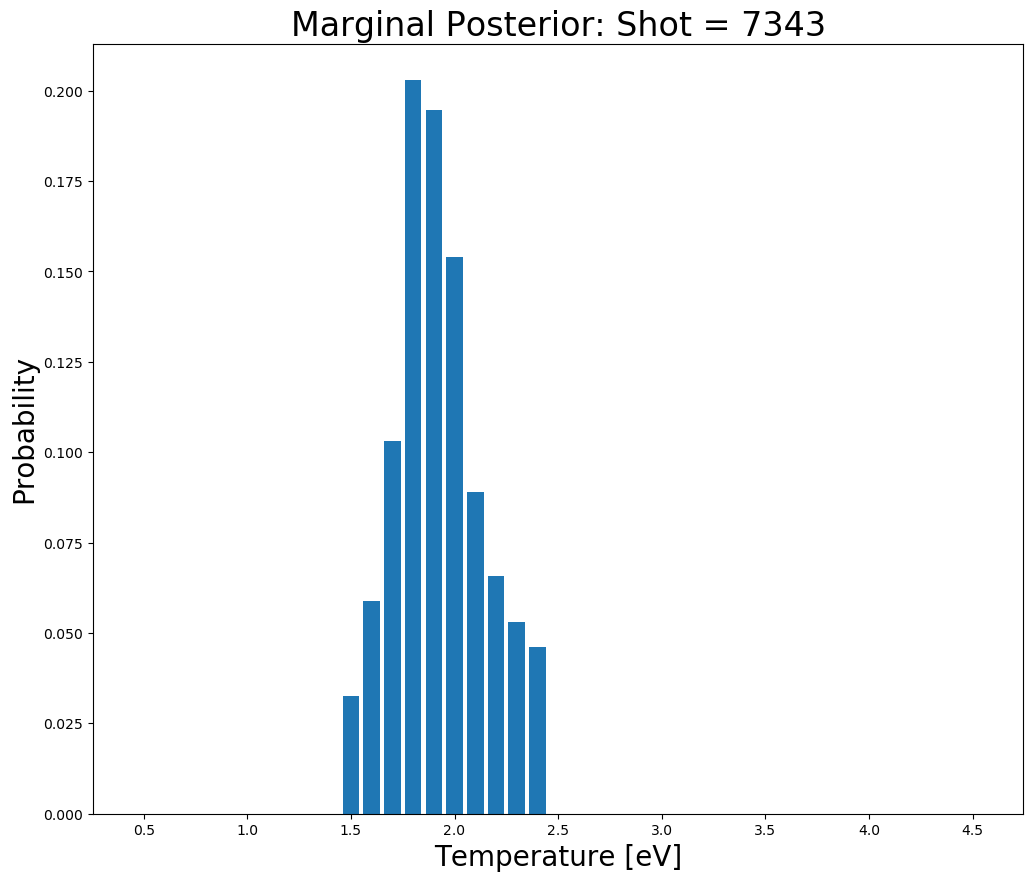}
\end{subfigure}
\hfill
\begin{subfigure}[t]{0.45\linewidth}
  \centering
  \includegraphics[width=\linewidth]{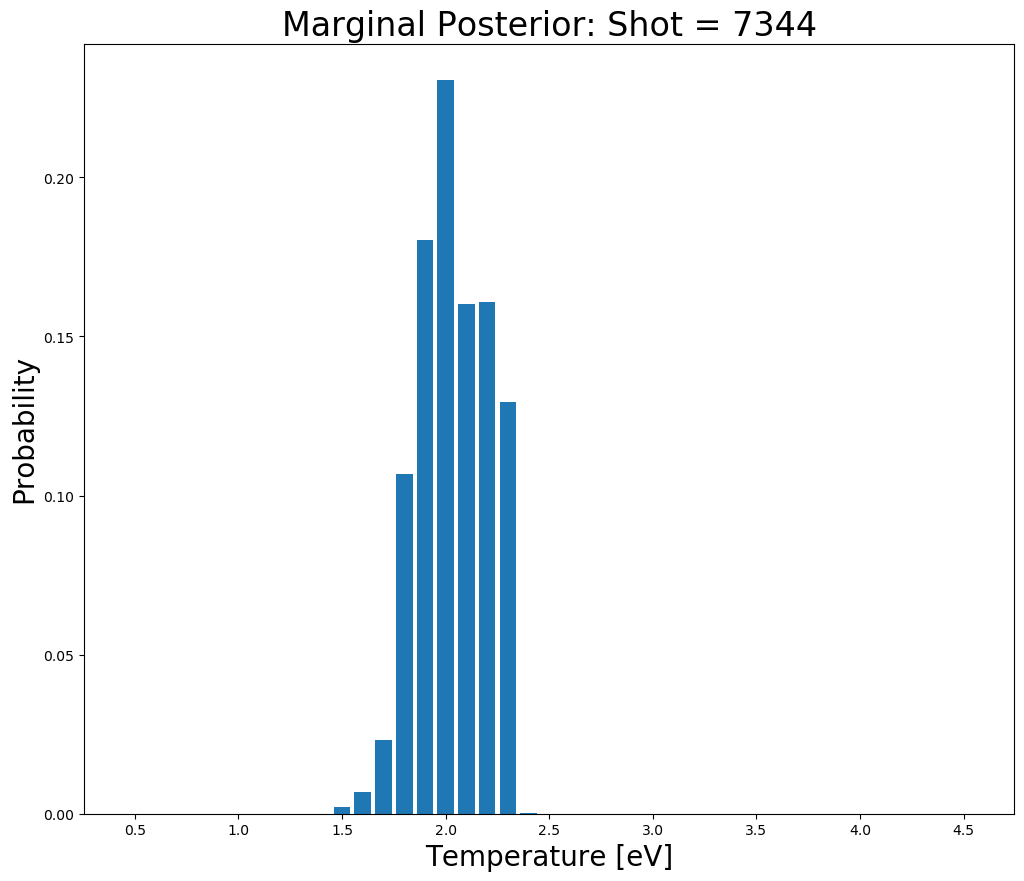}
\end{subfigure}
\end{center}
\caption{Marginal posterior probability distributions for temperature inferred from spectra recorded during six PLX shots on
  10/1/2021.}
\label{fig:sample-marginal-posteriors}
\end{figure}

\begin{figure}[htp]
\begin{center}
\includegraphics[width=0.9\linewidth]{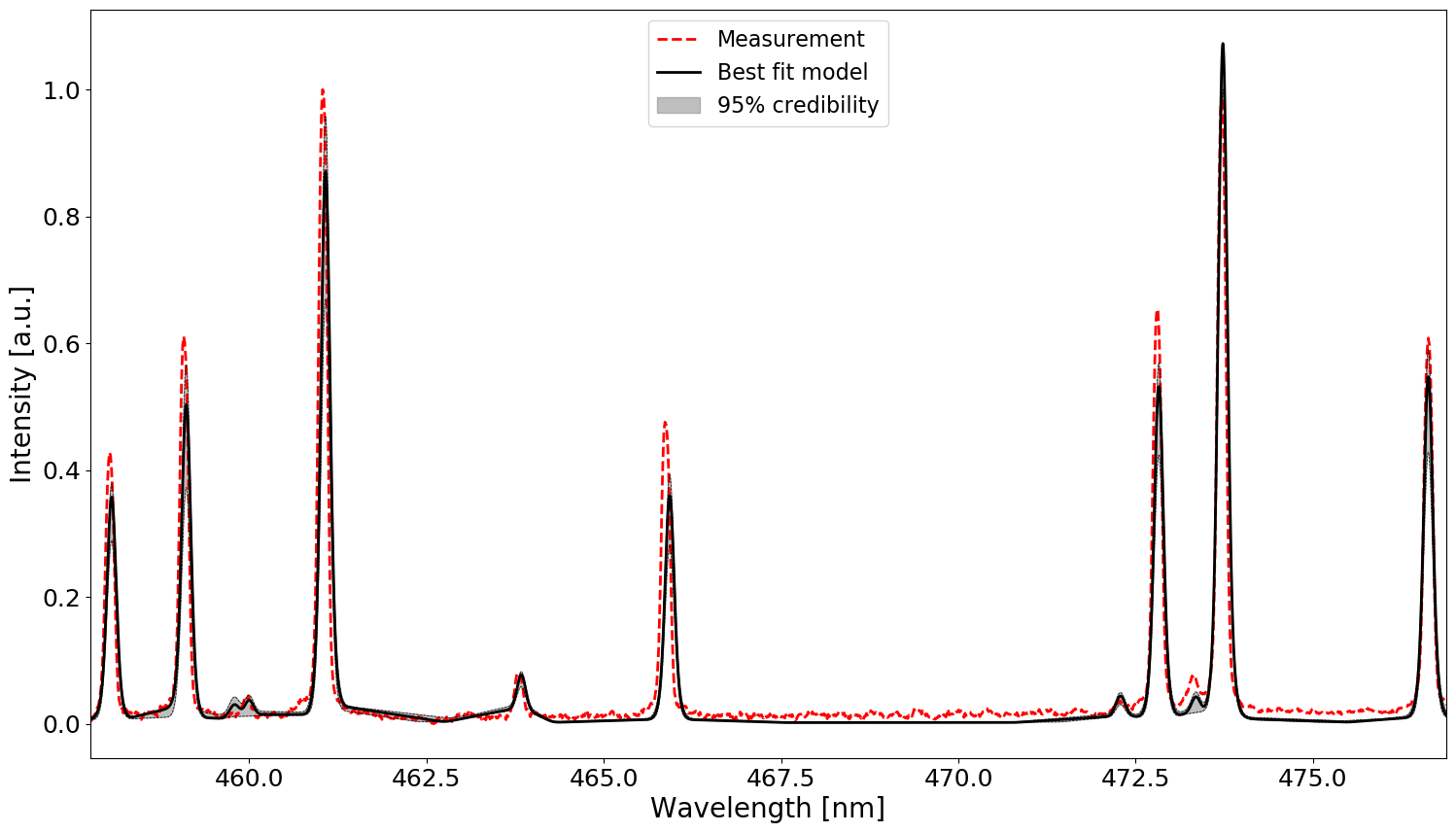}
\end{center}
\caption{Comparison of the measured and modeled spectrum for shot 7330.}
\label{fig:7330-spectrum}
\end{figure}

To get a better sense of the inferred temperature, since this is often
the primary quantity of interest of such measurements, the joint
posterior is marginalized over the density to give the marginal
posterior distribution for the temperature alone; specifically,
\begin{equation*}
\pi(T | I) = \int \pi (\rho, T | I) \, d\rho.
\end{equation*}
Figure~\ref{fig:sample-marginal-posteriors} shows these marginal
distributions for the same shots as in
Figure~\ref{fig:sample-joint-posteriors}.
Again, the marginal probability results demonstrate that the
uncertainties are signficant.  Qualitatively speaking, these result
from the fact that model results at multiple conditions provide
similar fits to the measured spectrum.  Figure~\ref{fig:7330-spectrum}
demonstrates this fact using shot 7330 as an example.  The figure
shows that over the range of highly probable conditions, as determined
by the inference process, the spectrum is qualitatively very similar.
Thus, given the inherent measurement precision and modeling errors, it
is difficult to determine the conditions more precisely given the
available information.

\subsubsection{Shot sequence results}\label{subsec:shots}
\begin{figure}[htp]
  \centering
  \includegraphics[width=0.6\linewidth]{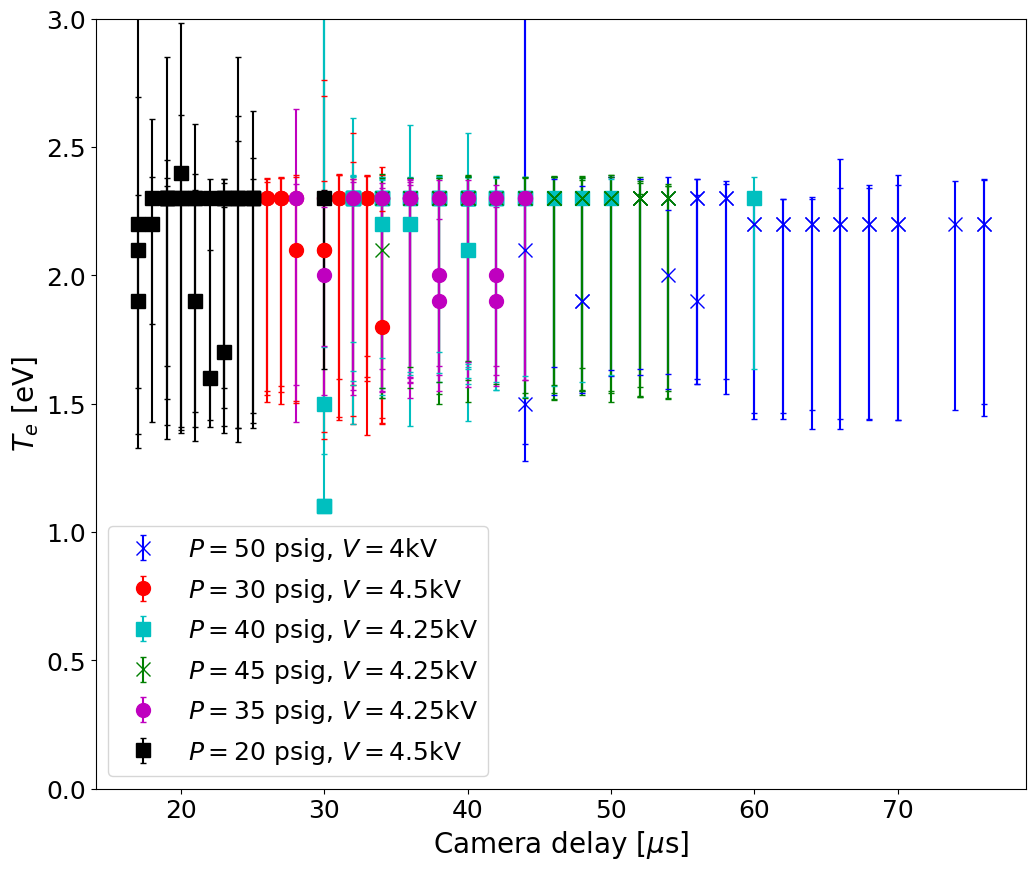}
\caption{Inferred temperature as a function of camera delay time for various
inlet pressures and capacitor bank volrages.  The symbols indicate the
maximum a posteriori temperature and the uncertainty bar indicates the
$95\%$ credibility interval for $T_e$.  Some conditions are repeated,
leading to different results at the same nominal conditions.}
\label{fig:T_vs_camera_delay}
\end{figure}

In addition to examining individual shots, the spectra obtained at
different conditions can be analyzed in an attempt to identify the
dependency of the state of the plasma upon various experimentally
controllable parameters.  As an example, Figure~\ref{fig:T_vs_camera_delay} shows the
dependence of the inferred temperature on the camera delay time, which
corresponds to the time in the evolution of the plasma formed in the
core of the merging jets, for the different inlet pressure and
capacitor bank voltages shown in Table~\ref{tab:exp_metadata}.  The results show that,
with the present measurements and models, it is very difficult to
identify any meaningful trends, because the uncertainty is large
enough that it is unclear whether changes in the most probable
temperature between conditions are real.
To evaluate the uncertainty estimates, we note that the data set
includes some conditions that are repeated.  These repeats lead to a
scatter in the results at the same delay times shown in the figure,
but all the repeats fall within the uncertainty of each other.  This
gives some confidence that the uncertain estimates from the inference
are not too small.  Further, the scatter between repeats is roughly of
the same magnitude as the uncertainty, indicating that our uncertainty
estimates are not dramatically too large.  Thus, we are forced to
conclude that any trends in the temperature are swamped by the
uncertainty and cannot be determined from these measurements alone.

\section{Conclusions}\label{sec:conclusion}
In this study, we have presented an approach for the automated estimation of plasma temperature and density from emission spectroscopy, integrating physical and probabilistic models by way of Bayesian inference. The robustness of our approach lies in its ability to assimilate complex physical models with probabilistic reasoning, offering a nuanced understanding of plasma characteristics. The detailed overview of Bayesian inference, coupled with the foundational physical models, set the stage for a methodological framework that is both rigorous and adaptable.

Our experimental results on a series of experiments at PLX have yielded promising results and consist of a pre-processing phase and a Bayesian inference phase. The results from this inference, particularly the individual shot and shot sequence analyses, have demonstrated the efficacy of our approach in providing accurate and reliable estimations of plasma parameters. These results underscore the potential of Bayesian methods in the field of plasma diagnostics, often offering a significant leap over traditional estimation techniques.

The current state of the technique offers many paths for possible extensions and improvements.  First, for practical reasons, the inferences of some parameters---e.g., the hyperparameters of the probabilistic model---are conducted separately from the primary inference for the plasma properties.  These inferences could be unified, allowing the effects of calibration uncertainties to be included and quantified.  Second, the process could be extended to enable inference for trace species, enabling the identification of contaminants.  More broadly, there is significant potential for adoption of these techniques in plasma research and other related fields. For instance, the ability to accurately estimate plasma temperature and density is not only critical for understanding plasma behavior but also pivotal for advancing nuclear instrumentation and its applications.

In conclusion, this study contributes a novel and effective tool for the field of plasma diagnostics, harnessing the power of Bayesian inference and sophisticated physical modeling. It opens up new avenues for research and application, promising to enhance the precision and understanding of plasma processes, as well as finding application in other areas of experimental analysis, such as analytical chemistry, etc. Future work will focus on refining the models further and exploring the integration of additional data sources, to extend the capabilities and applicability of the methodology.

\section{Acknowledgements}

This work was partially supported by U.S. ARPA-E Award No.~DE-AR0001265.   This manuscript has been authored in collaboration with Los Alamos National Laboratory/Triad National Security, LLC, Contract No. 89233218CNA000001, with the U.S. Department of Energy/National Nuclear Security Administration. The views and opinions of authors expressed herein do not necessarily state or reflect those of the United States Government or any agency thereof.

\bibliography{references}

\end{document}